\documentstyle[prl,aps,epsbox]{revtex}
\begin{document}
\draft
\preprint{}
\twocolumn[\hsize\textwidth\columnwidth\hsize\csname@twocolumnfalse%
\endcsname
%
\title{
Resonance Tunneling in Double-Well Billiards
with a Pointlike Scatterer
}
\author{
Taksu Cheon
}
\address{
Department of Physics, Hosei University, Fujimi, 
Chiyoda-ku, Tokyo 102, Japan
}
\author{
T. Shigehara
}
\address{
Computer Centre, University of Tokyo, Yayoi, 
Bunkyo-ku, Tokyo 113, Japan 
}
\date{December 2, 1996}
%
\maketitle
\begin{abstract}
The coherent tunneling phenomenon is investigated in rectangular
billiards divided into two domains by a classically unclimbable
potential barrier.
We show that by placing a pointlike scatterer inside the
billiard, we can control the occurrence and the rate of 
the resonance tunneling.
The key role of the avoided crossing 
is stressed.

\end{abstract}
%
]
\narrowtext

The tunneling is a quintessentially quantum phenomenon.
However, with the path-integral formulation, a systematic
semiclassical approximation can be formulated, 
and the problem can be treated in terms of the instanton, or the
``classical orbit with imaginary time'' in the region classically
inaccessible \cite{CO85}.  One can assume, therefore, that the
dynamics of this virtual classical orbits largely determines the
characteristics of the tunneling phenomena. One can further conjecture
that the tunneling in non-integrable system, with its chaotic virtual
orbit, might display novel features unknown in the textbook examples 
of one-dimensional barrier penetration, or more generally,
the tunneling in integrable systems \cite{WI86,CR94}. 
Recent studies on
several model systems show this is indeed the case.  It has been
revealed that very slight variation of the shape of the barrier can
cause an unexpectedly large enhancement of the tunneling rate
\cite{BT90,LB90}.   
It is also found that the tunneling can be made arbitrary small 
by the proper choice of the shape parameter \cite{GD91}. 
Certain evidence seems to link these characteristics 
to the chaotic character of the
virtual classical trajectories \cite{BT90,TU94,SI95,CW96}.
At this point, however, there is no clear understanding
about when and how the large variation in the tunneling rate occurs
in non-integrable systems, and how this might be related to the chaos.

In this Letter, we start from the premise that we are still in need
of a model which is rich enough to display all the essential features
of the non-integrable tunneling, yet simple enough to allow the
physical intuition to the phenomena.  
As a candidate for such a model, we propose a two-dimensional 
quantum billiard in which a particle moves around in a rectangular 
domain which is divided into two sub-domains by a finite-height 
barrier of rectangular shape parallel to the outer boundary.
We further place a pointlike scatterer, or a Dirac's delta potential
in the domain on and off the barrier.  We look at the change
in the tunneling between two sub-domains while varying the location
of the delta scatterer.  This is intended as a ``minimal model''
for the change of the barrier shape.
We show in the following that this model enables us to identify the
key role of the diabolical points and avoided level crossing in the 
non-integrable tunneling. It also reveals the condition 
for the occurrence of the tunneling and its enhancement and suppression.
The mirror symmetry of the system is shown {\it not} to be a necessary
condition for the resonance tunneling.

%
\begin{figure}
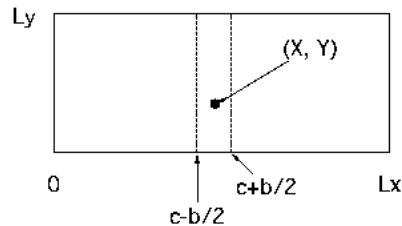

\center\psbox[hscale=0.5,vscale=0.5]{tnfig1.epsf}
\caption{
A schematic depiction of the model billiard system.  In the region
$c-b/2 < x < c+b/2$, a potential barrier of height $U_0$ is placed.
The pointlike scatterer located at $(X,Y)$ is indicated by 
a filled circle. 
}
\end{figure}
We consider a quantum particle moving inside a rectangular
hard wall of size $L_x \times L_y$ whose two sides are placed on the $x$
and $y$ axes.  We put a barrier potential of constant height $U_0$ 
\begin{eqnarray}
\label{1}
{U(\vec r)=U_0\theta (x_2-x)\theta (x-x_1)}
\end{eqnarray}
which divides the billiard into the left ($x < x_1$) and the right
($x > x_2$) sub-domains.  We call the center and the width 
of the barrier $c$ and $b$, so that we have $x_1 = c-b/2$
and $x_2 = c+b/2$.
We then place a delta potential of strength $v$ at $\vec r = \vec R$.
The model is depicted in Fig. 1.
Formally, the eigenvalue equation is
\begin{eqnarray}
\label{2}
\left[ {-{1 \over {2m}}\nabla ^2+U(\vec r)+v\delta (\vec r-\vec R)} 
\right]\psi _\alpha ({\vec r})
=\varepsilon _\alpha \psi _\alpha ({\vec r}). 
\end{eqnarray}
It is known that this equation is meaningless as it stands \cite{AG88}.
However, the problem is known to be
renormalizable with the introduction of the {\em formal} coupling
$\bar v$ in place of bare coupling $v$ \cite{AG88,SE90,SH94,SC96}.
We write the eigenvalues and eigenstates for $\bar v = 0$ as
$\eta_n$ and $\phi_n$, which we refer to as unperturbed solutions.
The full solution for $\bar v \ne 0$ is given in terms of 
the unperturbed basis.
There are two kinds of solutions.  The first, a trivial type 
of solution is
\begin{eqnarray}
\label{3}
\varepsilon_\alpha = \eta_n  \ \ \ \   {\rm if} \ \ \ \
\phi_n(\vec R) = 0,
\end{eqnarray}
that is, if $\vec R$ is on the node-line of the wavefunction
$\phi_n(\vec r)$.  
The second, more generic type of solution of eq. (\ref{2}) 
is obtained from the equation
\begin{eqnarray}
\label{4}
\overline G(\vec R;\varepsilon )-{1 \over {\bar v}}=0
\end{eqnarray}
where 
\begin{eqnarray}
\label{5}
\overline G(\vec r;\varepsilon )=\sum\limits_n {
  \phi_n(\vec r)^2
  \left[ {{1 \over {\varepsilon -\eta _n}}
    +{{\eta _n} \over {\eta _n^2+1}}} 
  \right] }
\end{eqnarray}
is the renormalized Green's function.
The eigenfunction corresponding to the eigenvalue 
$\varepsilon_\alpha$ is given by
\begin{eqnarray}
\label{6}
\psi_\alpha(\vec r)
=\sum\limits_n {
   {{\phi_n(\vec R)} \over {\varepsilon_\alpha -\eta_n}}
     \phi_n(\vec r) 
                }.
\end{eqnarray}

We first consider the case in which the system has mirror symmetry
with respect to the $x = L_x/2$ line which divides the billiard evenly
to left and right domains.  For this, we place the barrier at the center,
$c = L_x/2$, and also locate the pointlike scatterer 
along $x = L_x/2$ line, namely $X = L_x/2$.
Because of the mirror symmetry, the eigenfunction of the system possesses 
definite parities.  When $U_0$ is sufficiently large, there
are nearly degenerate parity doublets 
($\psi_{\alpha-}$, $\psi_{\alpha+}$) $\equiv$
($\psi_\alpha$, $\psi_{\alpha+1}$) with energies 
($\varepsilon_{\alpha-}$, $\varepsilon_{\alpha+}$) $\equiv$
($\varepsilon_\alpha$, $\varepsilon_{\alpha+1}$)
where we assume $\alpha$ to be an
odd integer.
From these parity doublets, one can construct the wave functions
\begin{eqnarray}
\label{7}
\chi_\alpha^{(\pm)} \approx {1 \over \sqrt{2}}
               \left( {\psi_{\alpha-} \pm \psi_{\alpha+}} \right)
\end{eqnarray}
which are localized to opposite sub-domains.
The localized states in each sub-domain $\chi_\alpha^{(\pm)}$ 
evolve according to
\begin{eqnarray}
\label{8}
e^{ i H t}\chi_\alpha^{(\pm)}
\approx \exp{(i\bar \varepsilon_\alpha t)}
\left( {\chi_\alpha^{(\pm)}\cos(\frac{\Delta_\alpha t}{2})
     - i\chi_\alpha^{(\mp)}\sin(\frac{\Delta_\alpha t}{2})}
\right)
\end{eqnarray}
where
\begin{eqnarray}
\label{9}
\bar \varepsilon_\alpha 
\equiv {{\varepsilon_{\alpha+}+\varepsilon_{\alpha-}} \over {2}}
\ \ \ \ \ {\rm and} \ \ \ \ \
\Delta_\alpha \equiv \varepsilon_{\alpha+}-\varepsilon_{\alpha-}
\end{eqnarray}
are the average energy and the energy splitting of the parity
doublet $\psi_{\alpha-}$ and $\psi_{\alpha+}$.
This means that the system oscillates between the states 
$\chi_\alpha^{(+)}$
and $\chi_\alpha^{(-)}$ with the half period $T = \pi/\Delta_\alpha$.
This is a quantum beating, or a resonance tunneling between two
nearly degenerate states.
The rate of the tunneling between the left and the right
domains $D_\alpha$ is given by
\begin{eqnarray}
\label{10}
D_\alpha \equiv {1 \over T} = {1 \over \pi} \Delta_\alpha.
\end{eqnarray}
Therefore, the energy splitting of the parity doublet 
$\Delta_\alpha$ is a direct 
measure of the tunneling rate.

\begin{figure}
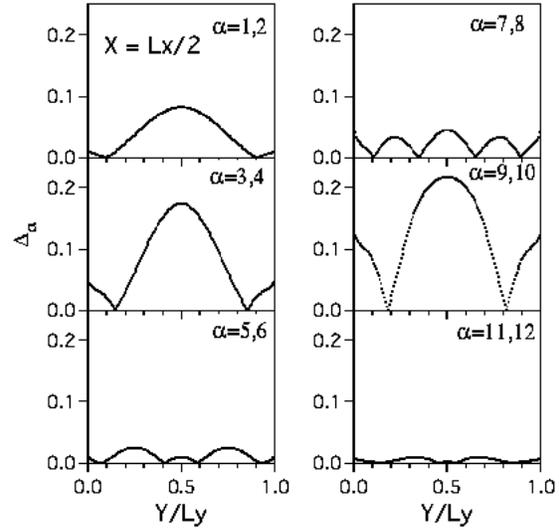

\center\psbox[hscale=0.5,vscale=0.5]{tnfig2.epsf}
\caption{
The energy splitting of the parity doublets as the function of 
perpendicular location of the pointlike scatterer.  Its horizontal
location is kept at the middle $X = L_x/2$.  The parameters are 
chosen as $L_x = 1.618$, $L_y = 1/L_x$, $U_0 = 50$, $c = L_x/2$,
$b = L_x/10$ and $\bar v = 100$.
}
\end{figure}
%
Because of the two dimensional nature of our model system,
there is still a freedom to choose the perpendicular location of
the pointlike scatterer, $Y$.  We look at the $Y$-dependence of
the energy splitting of the parity doublets.  In Fig. 2,
a specific example of $L_x=(\sqrt{5}+1)/2$,
$L_y=1/L_x$, $b=L_x/10$ and $U_0 = 50$ is shown. The mass of the system
is chosen to be $m = 2 \pi$, and the formal coupling is set to
$\bar v = 100$. With this choice of $\bar v$, 
the states near the ground state come into 
the strong coupling region \cite{SC96}. 
Most notable feature in the Figure is the fact that 
at certain values of $Y$, the energy splitting, thus the tunneling rate
is enhanced by several times compared to the unperturbed case,
which, in the Figure, is seen as $Y = 0$ and $Y = L_y$. 
Another point, no less important, is that there are several values of
$Y$ in which the energy splitting becomes zero.
This means that even with finite height potential,
one can totally suppress the tunneling by positioning the 
pointlike scatterer appropriately.
Since the Green's function $\overline G(\vec R, \varepsilon)$, eq. (5)
is a monotonously decreasing function of $\varepsilon$ except at
its poles, eq. (4) cannot have degenerate solutions.
The zero splitting of two energy eigenstates can occur 
only as the joint solution of eqs. (3) and (4).
Therefore, the location of the pointlike scatterer 
$\vec R^*$ at which the 
total suppression of the resonance tunneling occurs is determined by
\cite{CS96}
\begin{eqnarray}
\label{11}
\phi_n(\vec R^*) 
= \overline G(\vec R^*, \eta_n) - {1 \over \bar v} = 0.
\end{eqnarray}
Since $\vec R$ is a two dimensional vector, the solution of eq. (11)
are the isolated points.  These points are known as the 
{\em diabolical points} which play central role in the phenomena 
of Berry phase \cite{BE84}.  

\begin{figure}
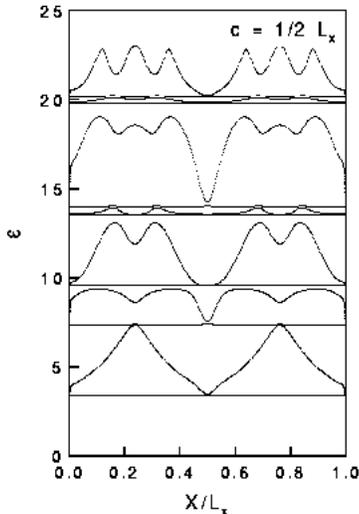

\center\psbox[hscale=0.5,vscale=0.5]{tnfig3.epsf}
\caption{
First fourteen eigenvalues as the function of horizontal location of 
the pointlike scatterer.  The perpendicular location is fixed at 
$Y = 0.4142 L_y$.  See Fig. 2 for the values of parameters.
The values $X = 0$ and  $L_x$ correspond to no pointlike scatterer.
}
\end{figure}
Next, we let the horizontal location of the pointlike scatterer 
$X$ to vary.  In Fig. 3, we show the first fourteen eigenvalues
as a function of X while fixing the $Y$ at $(\sqrt{2}-1) L_y$.  
One observes that two levels approach to each other not only
at $X =L_x /2$, but also at other places.  Indeed,
The solution of eq. (11) needs not to be 
restricted to the line $x = L_x/2$:  It can be on any nodeline 
$\phi_n(\vec r) = 0$ of the relevant unperturbed wave function.  
Examples are shown in Fig. 4, where we mark the diabolical locations
with asterisks for several eigenstates.
When the pointlike scatterer is at the diabolical locations, any 
combination of degenerate states can be chosen as the eigenstate.
One can form two eigenstates which cross over the barrier to have
equal amount of probability distribution in left and right
sub-domains with opposite relative signs at one sub-domain.
This is a generalization of the parity doublet states for the
mirror symmetric system.
The doublet states with equally partitioned wave function remain 
to be the eigenfunction of the system if the pointlike scatterer
is moved off the diabolical location in the direction
parallel to the y-axis along the nodeline $\phi_n({\vec r}) = 0$.
This is understood if one notices that one of the unperturbed
states remains to be the full eigenstate.
When the pointlike scatterer is not on the nodeline,
the eigenstates are localized either to the left or the right
sub-domains.  
The numerical example in Fig. 4, which depicts the parameter space
$(X, Y)$, illustrates the situation. 
When the pointlike scatterer is placed at shaded region, the
$\alpha$-th eigenstate is localized at the right sub-domain, and
when at unshaded region, localized at the left.  The figure is
drawn by defining an index 
\begin{eqnarray}
\label{12}
l_\alpha = \int_0^{x_d}{dx}\int_0^{L_y} 
                  {dy \psi^*_\alpha(x,y) \psi_\alpha(x,y)}
\end{eqnarray}
with $x_d$ signifying the location of the central nodeline of ``odd''
unperturbed wavefunction.  Typically, $x_d$ is close to the 
center of the barrier, namely, $x_d \simeq c$.
The shade indicates the region
 where $l_\alpha$ exceeds $1/2$.
It is on the line of $l_\alpha = 1/2$ which separates the shaded 
and unshaded regions where the resonance tunneling can occur.
This example clearly shows that it is not the mirror symmetry of the 
problem that brings the resonance tunneling phenomenon, but rather
the existence of the crossover between the two wave
functions with opposite localization characteristics \cite{PE91}.

\begin{figure}
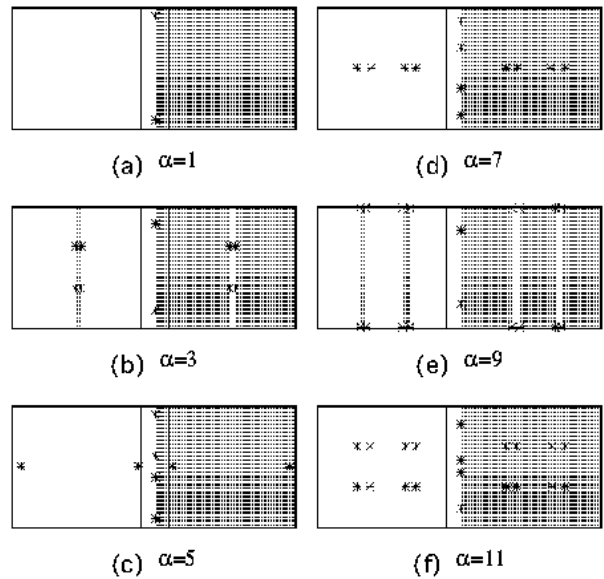

\center\psbox[hscale=0.5,vscale=0.5]{tnfig4.epsf}
\caption{
This chart shows the parameter space $(X,Y)$ with
the localization status of the $\alpha$-th eigen function. 
The shaded area represents the location of 
the pointlike scatterer
at which the localization index $l_\alpha$ is more than 1/2. 
The asterisks are the diabolical locations. 
The resonance tunneling occurs near the interface of 
shaded and unshaded areas.
The barrier is placed at the middle ($c=L_x/2$). 
Other parameters are also chosen as in Fig. 2.
}
\end{figure}
We now consider the more generic case in which the mirror 
symmetry is broken in the unperturbed level.  
As an example, we place the barrier of height $U_0 = 50$ and
width $b = 1/10 \times L_x$ at slightly left of the middle 
$c = 19/40 \times L_x$.  We thus have
$x_1 = 17/40 \times L_x$ and $x_2 = 21/40 \times L_x$.  
Other parameters are kept to be same as before.
The unperturbed states again form the doublets 
\{$\phi_\alpha$, $\phi_{\alpha+1}$\} ($\alpha$: odd).  
But the states $\phi_\alpha$ and $\phi_{\alpha+1}$ are
localized in the opposite side of the barrier.  Therefore,
no resonance tunneling can be observed in the unperturbed system.
We proceed to place the pointlike scatterer of strength 
$\bar v = 100$ at the location $(X, Y)$.
In Fig. 5, we show the first fourteen eigenvalues as a function
of $X$ with a fixed value for $Y$.
($Y = (\sqrt{2}-1)L_y$ in this example.)  One can spot several
avoided level crossings.  Specifically, the ones near 
$X = L_x/2$ appear to separate the two distinct regions of the
parameter space.
\begin{figure}
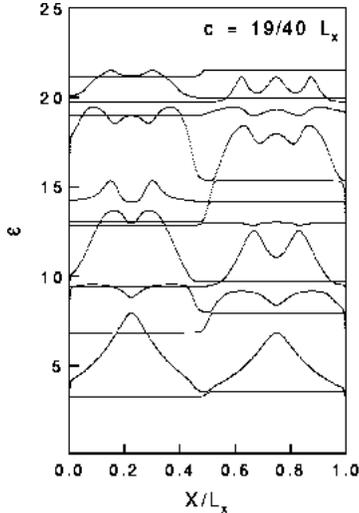

\center\psbox[hscale=0.5,vscale=0.5]{tnfig5.epsf}
\caption{
Same as Fig. 3 except that the barrier is slightly shifted leftward
($c = 19/40 L_x$).
}
\end{figure}
%
\begin{figure}
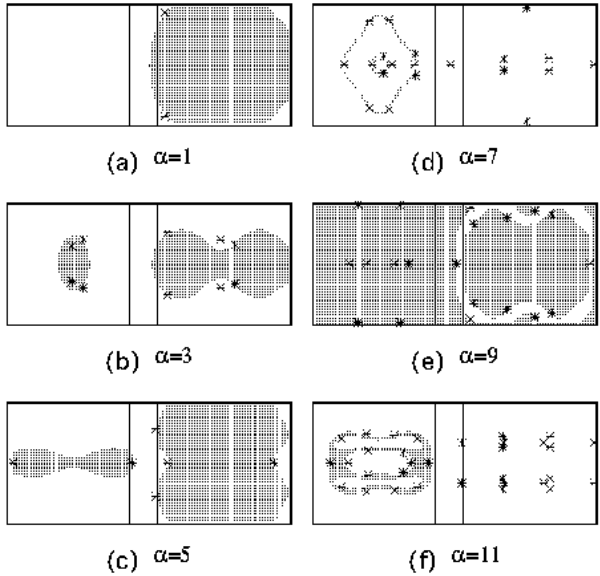

\center\psbox[hscale=0.5,vscale=0.5]{tnfig6.epsf}
\caption{
Same as Fig. 4 except that the barrier is slightly shifted leftward
($c = 19/40 L_x$).
}
\end{figure}
In Fig. 6, the parameter space $(X, Y)$
is depicted with diabolical points (asterisks) and localization
characteristics (shades) in an analogous manner as in Fig. 4.  
Namely, when the pointlike scatterer is placed at shaded region, the
$\alpha$-th eigenstate is localized at the right sub-domain, and
when at unshaded region, localized at the left.
The line of resonance 
tunneling on which $l_\alpha = 1/2$ now becomes the closed loops 
which separate the regions into left and right localization.
The resonance tunneling can occur only when $\vec R$ is located very 
close to this loop.
When one moves the pointlike scatterer along the loop, the rate
of the tunneling goes through a variation.
The rate becomes zero at the diabolical points,
and the maximum values tends to come at the points on the loop
in between the diabolical points in a manner similar to 
the one found in Fig. 2.  We conclude that the mirror symmetry
is not essential to obtain the resonance tunneling in double-well
billiards.

The mechanism behind the localization structure shown in Fig. 6
can be further elucidated by a closer inspection 
of the energy level diagram around the avoided crossing region.
There, we can assume that considering the mixing of two levels is
sufficient.
For the unperturbed doublet $\{ \phi_{\alpha},\phi_{\alpha+1}\}$
of opposite localization properties,
we can approximate eqs.(4) and (5) as
\begin{eqnarray}
\label{13}
{ {\phi_{\alpha}(\vec{R})^{2}} 
\over {\varepsilon-\eta_{\alpha}} } +
{ {\phi_{\alpha+1}(\vec{R})^{2}} 
\over {\varepsilon-\eta_{\alpha+1}} } = \frac{1}{v_{e\!f\!\!f}} 
\end{eqnarray}
where we define
\begin{eqnarray}
\label{14}
\frac{1}{v_{e\!f\!\!f}} \equiv \frac{1}{\bar{v}}
&-&\sum_{n \neq \alpha, \alpha+1} 
   \phi_{n}(\vec{R})^{2}
  \left(
  \frac{1}{\frac{(\eta_{\alpha}+\eta_{\alpha+1})}{2}-\eta_{n}}
  +\frac{\eta_{n}}{\eta_{n}^{2}+1} 
  \right)  \\ \nonumber
&-& \phi_{\alpha}(\vec{R})^{2}
  \frac{\eta_{\alpha}}{\eta_{\alpha}^{2}+1} -
  \phi_{\alpha+1}(\vec{R})^{2}
  \frac{\eta_{\alpha+1}}{\eta_{\alpha+1}^{2}+1}.
\end{eqnarray}
Here, we have replaced the energy $\varepsilon$ in all terms other than
$n = \alpha$ and $n =\alpha+1$ by $(\eta_{\alpha}+\eta_{\alpha+1})/2$.
The quantity $v_{e\!f\!\!f}$ is a function of both $\alpha$ and $\vec R$.
The solution of eq. (13) is readily obtained as
\begin{eqnarray}
\label{15}
\varepsilon_{\alpha\pm} 
&=&  \frac{(\eta_{\alpha}+\eta_{\alpha+1})}{2}+
\frac{v_{e\!f\!\!f}(f_{\alpha}+f_{\alpha+1})}
{2}    \\ \nonumber
&\pm& \frac{1}{2} 
\sqrt{ \{ \eta_{\alpha+1}-\eta_{\alpha} 
+v_{e\!f\!\!f}(f_{\alpha+1}-f_\alpha) \}^{2} 
+ 4 v_{e\!f\!\!f}^{2} f_{\alpha} f_{\alpha+1} 
     }.
\end{eqnarray}
where we have used the notation $f_\alpha \equiv \phi_\alpha(\vec R)^2$.
The corresponding eigenfunctions are given, 
apart from the normalization, by
\begin{eqnarray}
\label{16}
\psi_{\alpha\pm}(\vec{r}) \propto 
 \frac{\phi_{\alpha}(\vec{R})}
{\varepsilon_{\alpha\pm}-\eta_{\alpha}}
 \phi_{\alpha}(\vec{r})
+ \frac{\phi_{\alpha+1}(\vec{R})}
{\varepsilon_{\alpha\pm}-\eta_{\alpha+1}}
 \phi_{\alpha+1}(\vec{r}).
\end{eqnarray}
Along a path of varying $\vec R$ (for example, along the line
$Y$ = constant as in Fig. 5),
the avoided crossing takes place at the minimum value of the 
energy splitting $|\varepsilon_{\alpha+} -\varepsilon_{\alpha-}|$. 
This occurs when we have
\begin{mathletters}
\label{17}
\begin{eqnarray}
\label{17a}
\phi_{\alpha}(\vec{R})^{2} \gg \phi_{\alpha+1}(\vec{R})^{2}
\ \ \  {\rm and} \ \ \ 
\frac{1}{v_{e\!f\!\!f}} = \frac{\phi_{\alpha}(\vec{R})^{2}}
{\eta_{\alpha+1}-\eta_{\alpha}} 
\end{eqnarray}
for the case of $v_{e\!f\!\!f} > 0$, and when
\begin{eqnarray}
\label{17b}
\phi_{\alpha}(\vec{R})^{2} \ll \phi_{\alpha+1}(\vec{R})^{2} 
\ \ \  {\rm and} \ \ \
\frac{1}{v_{e\!f\!\!f}} = - \frac{\phi_{\alpha+1}(\vec{R})^{2}}
{\eta_{\alpha+1}-\eta_{\alpha}}
\end{eqnarray}
for $v_{e\!f\!\!f} < 0$.
\end{mathletters}
We consider the case of (a). Analogous argument can be
made for  case (b).  Inserting the second condition of eq.(17a) 
into eq. (15), we obtain 
\begin{eqnarray}
\label{18}
\varepsilon_{\alpha\pm} = \frac{\eta_{\alpha}+\eta_{\alpha+1}}{2}+
\frac{\eta_{\alpha+1}-\eta_{\alpha}}{2} 
\left( 1+\gamma^{2} \pm \gamma \sqrt{4+\gamma^{2}} \right)
\end{eqnarray}
where we use the notation
$\gamma \equiv |\phi_{\alpha+1}(\vec R)/\phi_{\alpha}(\vec R)|$.
The ratio between the two components of $\psi_{\alpha\pm}(\vec{r})$ 
in eq. (16) is given by
\begin{eqnarray}
\label{19}
\left|
\frac{\phi_{\alpha+1}(\vec{R})}
{\varepsilon_{\alpha\pm}-\eta_{\alpha+1}}
/ 
\frac{\phi_{\alpha}(\vec{R})}
{\varepsilon_{\alpha\pm}-\eta_{\alpha}} 
\right|
= \left| \frac{2}{\gamma \pm \sqrt{4+\gamma^{2}}} + \gamma
  \right| 
\end{eqnarray}
Because we have $\gamma \simeq 0$ (the first condition of eq. (17a)),
this ratio becomes very close to 1, 
giving one-to-one mixing of the
states of opposite localization properties.
We can conclude that the resonance tunneling occurs
at the parameter values corresponding to the avoided crossing.
Therefore, in the whole parameter space $(X, Y)$,
the line of resonance tunneling
is the ``valley line'' of minima of the energy
splitting $|\varepsilon_{\alpha+} - \varepsilon_{\alpha-}|$.
At the diabolical points, $\gamma$ becomes exactly zero.
We also observe from eq. (18), that both levels are 
close to the unperturbed level $\varepsilon = \eta_{\alpha+1}$ 
at the avoided crossing. (This becomes $\varepsilon = \eta_{\alpha}$
for $v_{e\!f\!\!f}< 0$ case.)
These features are exactly what one finds in the numerical
example of Figs. 5 and 6.

Finally, we discuss our results in a broader view.
The role of the diabolical points in the resonance tunneling has
been already recognized by the authors of refs. \cite{GD91} and 
\cite{TU94}.
The clarification of its distribution in the parameter space 
and its decisive role in shaping the line of resonance tunneling is
made possible only with the very simple setting
of the current work.
Although the model studied here is a simplistic one, we can conjecture 
that our main finding is applicable to generic non-integrable tunneling
systems:
The resonance tunneling occurs along the ``valley line'' of avoided
crossing in the parameter space of the barrier shape, and the rate of
tunneling undergoes a large variation when the shape parameters go 
through the diabolical points. 

In the current study, we have made no immediate
reference to chaotic motion nor to the semiclassical analysis,
since the system is solvable in full quantum mechanical setting.
However, if we look at the doublet states in higher energy
with higher number of nodes, we expect the appearance of 
complex pattern both in the distribution of the diabolical 
locations and in the shape of the line on which resonance 
tunneling takes place.
One can also make the tunneling more complex in above sense
by placing more than one pointlike scatterers in the billiards.
The semiclassical approach to such problem will be an
interesting and challenging task.

It is known that the effect of the pointlike scatterer disappears
in the classical limit \cite{CS96}.  The fact that the
classically non-observable object causes the large change in 
the tunneling rate is a good example showing the 
intricacy of the quantum physics.
We also stress that the system considered here is not just
of academic nature, but may actually be
tested either through quantum dot or through microwave experiments.
Even its application as a 
switching device in the nanometer scale might be speculated as
a remote possibility.

In summary, we have studied several characteristics of 
non-integrable tunneling phenomena through the numerical
examples of a double-well billiard with a pointlike
scatterer.  We have identified the key role of the
diabolical crossings in the tunneling.

We acknowledge the helpful discussion with
the members of the theory division of Institute
for Nuclear Study, University of Tokyo.
%
%
%

%
%
%
%
%

\begin{thebibliography}{10}
\bibitem{CO85}
S. Coleman, {\it Aspects of Symmetry}
(Cambridge University Press, Cambridge, 1985).  
\bibitem{WI86}
M Wilkinson, 
Physica {\bf D21} (1986) 341.
\bibitem{CR94}
S. C. Creagh,
J. Phys. {\bf A27} (1994) 4969.
\bibitem{BT90}
O. Bohigas, S. Tomsovic and D. Ullmo, 
Phys. Rev. Lett. {\bf 65} (1990) 5.
\bibitem{LB90}
W. A. Lin and L. E. Ballentine, 
Phys. Rev. Lett. {\bf 65} (1990) 2927.
\bibitem{GD91} 
F. Grossmann, T. Dittrich, P. Jung and P. H{\"a}nggi,
Phys. Rev. Lett. {\bf 67} (1991) 516.
\bibitem{TU94}
S. Tomsovic and D. Ullmo,
Phys. Rev. {\bf E50} (1994) 145.
\bibitem{SI95}
A. Shudo and K. S. Ikeda, 
Phys. Rev. Lett. {\bf 74} (1995) 682.
\bibitem{CW96}
S. C. Creagh and N. D. Whelan, LANL preprint chao-dyn/9608002.
\bibitem{AG88}
S. Albeverio, F. Gesztesy, R. H{\o}egh-Krohn and H. Holden, 
{\it Solvable Models in Quantum Mechanics}
(Springer-Verlag, New York,1988).
\bibitem{SE90}
P. {\v S}eba, Phys. Rev. Lett. {\bf 64} (1990) 1855.
\bibitem{SH94}
T. Shigehara, Phys. Rev. {\bf E50} (1994) 4357.
\bibitem{SC96}
T. Shigehara and T. Cheon, Phys. Rev. {\bf E54} (1996) 1321.
\bibitem{CS96}
T. Cheon and T. Shigehara, Phys. Rev. Lett. {\bf 76} (1996) 1770.
\bibitem{BE84}
M. V. Berry, Proc. Roy. Soc. Lond. {\bf A392} (1984) 43.
\bibitem{PE91}
A. Peres, Phys. Rev. Lett. {\bf 67} (1991) 158.
\bibitem{BE89}
M. V. Berry, Proc. Roy. Soc. London {\bf 422} (1989) 7.
\end{thebibliography}
\end{document}